\documentclass[aps,prd,twocolumn,showpacs]{revtex4} 

\usepackage{graphicx}

\usepackage{amsmath}
\usepackage{amssymb}

\font\FermiSmallfont=cmssq8 scaled 1200

\def\LANLppthead#1#2{
\null 
\begin{center}\vskip -1.0truein{\hbox to 7.5truein {
\hfill
\vbox to 1in {\vfill \FermiSmallfont
              \hbox{#1}
              \hbox{#2}
              \vfill}
}}\vskip-0.0truein\end{center}}

\begin{document}

\LANLppthead {LA-UR 05-8831}{astro-ph/0511630}

\title{Production and Evolution of Perturbations of Sterile Neutrino
  Dark Matter}

\author{Kevork Abazajian}
\affiliation{Theoretical Division, MS B285, Los Alamos National
  Laboratory, Los Alamos, NM 87545 }


\begin{abstract}
  Sterile neutrinos, fermions with no standard model couplings [$\rm
    SU(2)$ singlets], are predicted by most extensions of the standard
  model, and may be the dark matter.  I describe the nonthermal
  production and linear perturbation evolution in the early universe
  of this dark matter candidate.  I calculate production of sterile
  neutrino dark matter including effects of Friedmann dynamics
  dictated by the quark-hadron transition and particle population, the
  alteration of finite temperature effective mass of active neutrinos
  due to the presence of thermal leptons, and heating of the coupled
  species due to the disappearance of degrees of freedom in the
  plasma.  These effects leave the sterile neutrinos with a
  non-trivial momentum distribution.  I also calculate the evolution
  of sterile neutrino density perturbations in the early universe
  through the linear regime and provide a fitting function form for
  the transfer function describing the suppression of small scale
  fluctuations for this warm dark matter candidate.  The results
  presented here differ quantitatively from previous work due to the
  inclusion here of the relevant physical effects during the
  production epoch.
\end{abstract}

\maketitle

\section{Introduction}

The nature of dark matter remains one of the most significant unsolved
problems in cosmology and particle physics.  The abundance of dark
matter has been precisely determined by observations of anisotropies
in the cosmic microwave background~\cite{Spergel:2003cb}, and the
measurements of the growth of cosmological structure in the clustering
of galaxies~\cite{Tegmark:2003ud} and in the Lyman-$\alpha$
forest~\cite{Seljak:2004xh}.  The fundamental nature of the dark
matter, however, remains unknown.

One natural candidate is a fermion that has no standard model
interactions other than a coupling to the standard neutrinos through
their mass generation mechanism~\cite{Dodelson:1993je,Shi:1998km}.
Due to their lack of interactions and association with the neutrino
sector, such fermions are referred to as sterile neutrinos.
Observations are consistent with sterile neutrinos as the dark matter
for a narrow mass range for the standard production mechanism.  In
this allowed range of masses, the sterile neutrino has a
non-negligible thermal velocity component, and is therefore a warm
dark matter (WDM) candidate.

The prevalent ansatz of an absolute cold dark matter (CDM) component
in galaxy formation is not strictly valid even for one of the most cited CDM
candidate, the lightest supersymmetric particle, which has a small
but non-zero velocity dispersion
\cite{Hofmann:2001bi,Jungman:1995df}.  The damping scale at which
thermal velocities of the dark matter cut off the growth of
gravitationally bound structures remains an open question.  One
principal challenge to the CDM paradigm is the order of magnitude
over-prediction of the observed satellites in galaxy-sized halos such
as the Milky Way
\cite{Kauffmann:1993gv,Klypin:1999uc,Moore:1999wf,Willman:2004xc}.
Warm dark matter suppresses dwarf galaxy formation, which may occur
through fragmentation of larger structures~\cite{Bode:2000gq}.
Semi-analytic galaxy formation modeling has found that the number of
dwarf galaxies formed in satellite halos may be suppressed due to the
reionization, stellar feedback within halos, and/or tidal stripping of
satellites
\cite{Dekel:1986gu,Thoul,Bullock:2000qf,Bullock:2000wn,Benson:2001au}.
Such semi-analytic modeling is both powerful and malleable, and must
be verified in robust hydrodynamic simulations of galaxy formation.  
Whether a minor or major suppression of small mass halos is beneficial
or detrimental to the suppression of dwarf galaxy formation remains
unsolved.

Four more problems in the CDM paradigm may benefit from the reduction
of power on small scales from WDM. First is the reduction of the
prevalence of halos in low-density voids in N-body simulations of CDM
structure formation, consistent the apparent dearth of massive galaxies
within voids in local galaxy
surveys~\cite{Peebles:2001nv,Bode:2000gq}. The second is the
relatively low concentrations of galaxies observed in rotation curves
compared to what is predicted from the $\Lambda$CDM power
spectrum~\cite{Dalcanton:2000hn,vandenBosch:2000rz}, which can be
relieved by a reduction of the initial power spectrum of density
fluctuations at small
scales~\cite{Zentner:2002xt,Abazajian:2005kz}. The third is the
``angular-momentum'' problem of CDM halos, where gas cools at very
early times into small mass halos and leads to massive low-angular
momentum gas cores in galaxies, which can be alleviated by the
hindrance of gas collapse and angular momentum loss through the delay
of small halo formation in a WDM scenario~\cite{Dolgov:2001nq}.  The
fourth problem is the formation of disk-dominated or pure-disk
galaxies in CDM models, which is impeded by bulge formation due to the
high merger accretion rate history in CDM models, but may be
alleviated with WDM~\cite{Governato:2002cv,Kormendy2005}.

Sterile neutrinos arise naturally in most extensions to the standard
model of particle physics.  Singlet neutrinos with masses relevant to
oscillation experiments and dark matter can arise in grand unified
models~\cite{Brahmachari:1998kt}, string-inspired
models~\cite{Langacker:1998ut}, and models with large extra
dimensions~\cite{Dvali:1999cn,Abazajian:2000hw}.  In reverence to
Occam's razor, the minimalist model of Ref.~\cite{Asaka:2005an}
produces the neutrino-oscillation inferred neutrino mass hierarchy,
the dark matter via a sterile neutrino, as well the observed baryon
asymmetry in a neutrino minimal standard model ($\nu$MSM) that
introduces Majorana and Dirac neutrino mass terms to the standard
model Lagrangian.  There is also an indication that one or more light
sterile mass eigenstates may cause the flavor transformation seen in
the Los Alamos Liquid Scintillator Neutrino Detector (LSND)
experiment~\cite{Athanassopoulos:1997pv,Sorel:2003hf}.  Another
intriguing motivation for the presence of a dark matter sterile
neutrino is abundance of anomalously high pulsar velocities that may be
difficult to produce in the convective hydrodyamics in a supernova,
but may be produced in assymetric sterile neutrino emission from a hot
nascent neutron
star~\cite{Kusenko:1998bk,Fuller:2003gy,Kusenko:2004mm}.  However, whether
convective overturn and the resulting global asymmetry in the
ejecta alone can power the observed distribution of pulsar velocities
remains an open question~\cite{Scheck:2003rw,Fryer:2005sz}.

The potentially beneficial effects of the suppression of cosmological
small-scale structure in WDM can also lead to observational conflicts
if the suppression extends to excessively high mass and length scales.
As I shall show in detail below, the suppression scale monotonically
decreases with increasing sterile neutrino particle mass.  The
reionization of the universe by a redshift of $z\sim 6$ requires
sufficient structure formation at very early times, and can place one
of the most stringent lower-bounds on the sterile neutrino
mass~\cite{Barkana2001,Yoshida:2003rm}.  The radiative decay of the
sterile neutrino dark matter may, however, increase the hydrogen
ionization fraction, augmenting molecular hydrogen formation, gas
cooling, star formation, and therefore
reionization~\cite{Biermann:2006bu}.  
One of the best direct
measures of clustering at small scales is the clustering observed in
intervening gas along the line-of-sight to a quasar, known as the
Lyman-$\alpha$ forest~\cite{Narayanan:2000tp}.
Statistically-consistent constraints allowing freedom in all
cosmological parameters and constraints from the cosmic microwave
background, galaxy clustering, and a measurement of clustering in the
Lyman-$\alpha$ forest gives a lower limit for the sterile neutrino
dark matter particle mass as $m_s >1.7\rm\ keV \ (95\%\
CL)$~\cite{AbazajianLower05}.  

In all studies of the types discussed above of galaxy formation,
cosmological reionization and clustering in the Lyman-$\alpha$ forest,
robust conclusions require an accurate initial description of the
fluctuations arising from the early universe.  In \S\ref{production},
I outline the production mechanism of sterile neutrino dark matter
through the varying particle population in the early universe and the
QCD transition, and describe the resulting non-thermal sterile
neutrino energy distribution.  In \S\ref{perturbations}, I follow the
evolution of the non-thermal sterile neutrino dark matter
perturbations through the radiation dominated era into the linear
regime of the matter dominated era.  This is related to the CDM
perturbation spectrum through a transfer function.  My results differ
significantly from previous work on sterile neutrino dark matter
perturbation evolution~\cite{Hansen:2001zv,Viel:2005qj} which
neglected the effects of the changing particle population in the early
universe, the QCD transition, and the dilution of the dark matter due
to annihilation.  

The sterile neutrino particle dark matter candidate studied here may
be embedded in several extensions beyond the standard model of
particle physics, as a superpartner, or it may have properties that
would have it couple to other species at higher temperatures,
including the inflaton.  However, knowing its behavior in the early
universe in such extensions would involve introducing a much more
model dependent interaction and production mechanism, as well as
knowledge of all of the degrees of freedom that are present and may
annihilate and dilute the sterile neutrino in such a model between the
coupling epoch and today.  Even in such models with higher energy
scale couplings, dilution may render the abundance of sterile
neutrinos negligible upon entering the production epoch considered
here of $T < 300 {\rm\ MeV}$.  The production mechanism studied here
is a minimalist extension to the standard model through the neutrino
mass generation mechanism, and does not require what would be at this
point speculation of higher energy physics.  Therefore, the initial
abundance of sterile neutrinos entering the oscillation-production
epoch is taken here to be nil.

\section{Production}
\label{production}
Sterile neutrinos of interest for dark matter are never coupled to the
primordial plasma.  The production of sterile neutrinos in the early
universe within the mass range of interest for warm to cold dark
matter occurs at temperatures where collisions dominate the evolution
of the neutrino system, and matter-effected oscillations are
suppressed by induced thermal masses and the quantum zeno effect.  The
true time evolution of the system is described by that of the density
matrix~\cite{McKellar:1992ja}.  However, the collision dominated
regime allows a simplification of the density matrix evolution to a
quasi-classical Boltzmann equation of the form \cite{Abazajian:2001nj}
\begin{widetext}
\begin{equation}
\frac{\partial}{\partial{t}}f_s(p,t) - H\, p\,
\frac{\partial}{\partial{p}}f_s(p,t) \approx  \frac{1}{4}
\frac{\Gamma_\alpha(p) \Delta^2 (p) \sin^2 2\theta}{\Delta^2 (p) \sin^2
2\theta + D^2(p) + \left[\Delta (p) \cos 2\theta - V^L -
V^T(p)\right]^2} \left[f_\alpha(p,t) -
f_s(p,t)\right],
\label{fullboltz}
\end{equation}
\end{widetext}
with a corresponding evolution equation for the antineutrino
distributions.  Here, $f_s(p,t)$ and $f_\alpha(p,t)$ are the active
and sterile neutrino distribution functions, as a function of
momentum, $p$, and time, $t$; $H=\dot a/a$ is the Hubble expansion for
scale factor $a$;
$\Delta(p)= \delta m^2/2p$ is the vacuum oscillation factor dependent
on the mass-squared difference $\delta m^2 = m_s^2 - m_\alpha^2$
between the active and sterile neutrinos; the mixing angle between the
two flavor states is $\theta$.

The production is driven by the collision rate
\begin{equation}
\Gamma_\alpha (p) \approx \begin{cases}1.27\, G_{\rm F}^2 p T^4, \qquad
\alpha=e, \\ 0.92\, G_{\rm F}^2 p T^4,\qquad
\alpha=\mu,\tau, \end{cases}
\end{equation}
and is augmented at temperatures, $T$, above the quark-hadron (QCD)
transition, where quarks and the massive leptons $\mu$ and $\tau$
contribute~\cite{Abazajian:2001nj}, and are included in the
calculation here.  The system is damped by the quantum Zeno effect,
$D(p)=\Gamma_\alpha(p)/2$, and mixing is suppressed by the thermal
potential~\cite{Notzold:1987ik}
\begin{eqnarray}
\label{vt}
V^T (p) = &-& \frac{8\sqrt{2} G_{\rm F} p}{3 m_{\rm Z}^2}
\left(\langle E_{\nu_\alpha} \rangle n_{\nu_\alpha} + \langle
E_{\bar\nu_\alpha} \rangle n_{\bar\nu_\alpha}\right) \cr &-&
\frac{8\sqrt{2} G_{\rm F} p}{3 m_{\rm W}^2} \left(\langle E_\alpha
\rangle n_\alpha + \langle E_{\bar\alpha} \rangle
n_{\bar\alpha}\right),
\end{eqnarray}
which has contributions from thermally populated leptons of the same
flavor as the active neutrino.  

The asymmetric lepton potential for flavor $\alpha$ is
\begin{equation}
V^L = \sqrt{2}G_F\left[ 2\left(n_{\nu_\alpha}-n_{\bar\nu_\alpha}\right)
 +
 \sum_{\beta\neq\alpha}\left(n_{\nu_\beta}-n_{\bar\nu_\beta}\right) -\frac{n_n}{2}\right].
\end{equation}
Here, I will only consider lepton number symmetric universes;
however, the limits on the lepton asymmetry allow for a non-negligible
lepton number \cite{Dolgov:2002ab,Abazajian:2002qx,Wong:2002fa} which
can drive resonant sterile neutrino dark matter
production~\cite{Shi:1998km,Abazajian:2001nj}. One can neglect the
asymmetric potential due to the baryon number $n_n$, which is
subdominant at all temperature scales of interest for sterile neutrino
dark matter production.  The temperature of peak production is approximately
\begin{equation}
T_{\rm peak} \approx 130 {\rm\ MeV}\left(\frac{m_s}{3 {\rm\
keV}}\right)^{1/3}.\label{productiontemperature}
\end{equation}

The time evolution of the temperature-dependent thermal potential and
collision rate needed to integrate the production Boltzmann equation
requires knowing the time-temperature relation, and therefore the
evolution of the background plasma.  Details of the general
time-temperature relation are given in the appendix of
Ref.~\cite{Abazajian:2001nj}, and are summarized here.  The time
temperature relation depends on the expansion-dependent change of the
temperature
\begin{equation}
\frac{da}{dT} = \frac{d\rho_{\rm tot}}{dT} \left(\rho_{\rm tot} +
p_{\rm tot}\right)^{-1},
\label{drdtemp}
\end{equation}
and the Friedmann equation governing the expansion rate, $da/dt = 3H$.
The standard evolution of the pressure and temperature are affected by
the changing thermal population of particle species, which I calculate
using the known standard model particle mass
distribution~\cite{PDBook}.

A critical consideration for dark matter production is the fact that
production occurs near the temperature of the quark-hadron
transition~\cite{Abazajian:2002yz}.  Lattice QCD calculations of two
massless $u,d$ quarks and an infinitely massive $s$-quark show a
cross-over type transition at $T_{\rm QCD} = 173\pm
8\rm\ MeV$~\cite{Karsch:2000kv}; more realistic (2+1) quark flavor
lattice calculations find a transition at $T_{\rm QCD} = 169\pm
12\rm\ (stat.)\pm 4 (sys.)\ MeV$~\cite{Bernard:2004je}.  Overall, due
to hadron and lepton population as well as the quark-hadron
transition, the statistical degrees of freedom of the plasma, $g_*$,
change by nearly an order of magnitude in the temperature range of
interest for sterile neutrino dark matter production.  In addition,
the annihilation of these species heats the coupled plasma relative to
the (constantly) decoupled sterile neutrinos, which are subsequently
diluted and spectrally distorted.  Note that production is also
affected by the temperature range of the softness of the cross-over
transition.  Motivated by Ref.~\cite{Bernard:2004je}, I model the
softness of the crossover as rapid, over $5\rm\ MeV$ at $T_{\rm QCD} =
170\rm\ MeV$.  Changes to the softness of the transition affect the
production abundance by only a few percent.

For low-mass sterile neutrinos, $m_s < 0.5\rm\ keV$, production occurs
at low temperatures where the statistical degrees of freedom in the
primordial plasma are changing minimally, and one may set $g_*$ to be
constant so that the production Eq.~(\ref{fullboltz}) can be
integrated analytically~\cite{Dodelson:1993je}.  In this
approximation, the momentum dependence of the production disappears,
and the sterile neutrino momentum distribution is simply a suppressed active
neutrino distribution.

\begin{figure}
\includegraphics[width=3.3truein]{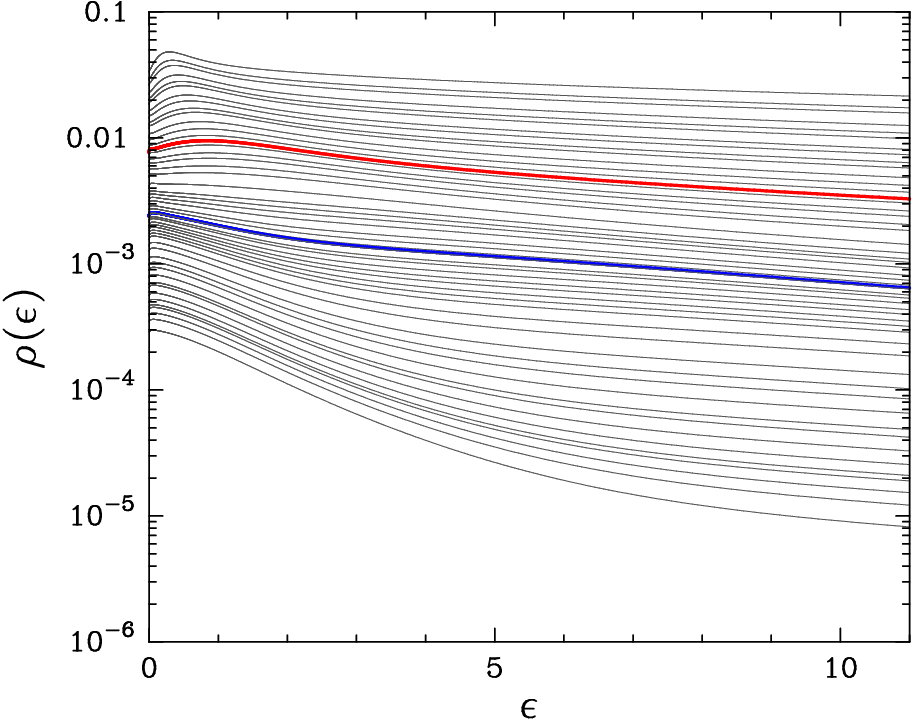}
\caption
{\small The resulting relative distribution coming from the production
  epoch for sterile neutrino dark matter relative to the active
  neutrinos, $\rho(\epsilon) = f_s(\epsilon)/f_\alpha(\epsilon)$,
  ($\epsilon \equiv p/T$) over a mass range $0.3 < m_s <
  140{\rm\ keV}$, for fifty cases, and increasing $m_s$ having
  decreasing distribution amplitudes.  All cases have $\Omega_{\rm DM} =
  0.26$. The upper thick (red) line is for the case of $m_s =
  1.7\rm\ keV$ and lower thick (blue) line is for the case $m_s =
  8.2\rm\ keV$. }
\label{sterile_dist}
\end{figure}

However, for $m_s \gtrsim 0.5\rm\ keV$, in order to accurately include
the strong changes in the plasma background, the quark-hadron
transition, dark matter dilution, the modification of scattering
rates, and flavor-dependent modification of the thermal potential
[Eq.~(\ref{vt})], the production equation must be integrated
numerically, and the result is momentum-dependent.  In order to test
the accuracy of the numerical calculation, I solve the analytically
solvable case of a simple power-law time-temperature relation as well
as all of the other simplifications of the analytic case, but with the
full momentum distribution of neutrinos.  The numerical integration
recovers the analytic result within 1\%.  In Fig.~\ref{sterile_dist},
I show the resulting relative distribution of sterile to active
neutrinos $\rho(\epsilon) = f_s(\epsilon)/f_\alpha(\epsilon)$ for
masses in the range $0.3 < m_s < 140{\rm\ keV}$, with
$\epsilon \equiv p/T$.  Note that all cases are distorted from a pure
constant suppression, and are increasingly distorted for more massive
neutrinos, which are produced at higher temperatures
[Eq.~(\ref{productiontemperature})] where the effects described above
are more pronounced.  One of the dominant effects is the cooling and
subsequent enhancement of the low momentum sterile neutrino
distribution due to the heating of the plasma, including the active
neutrinos, from particle annihilations.  Another feature is the
enhancement at $\epsilon < 1$ for lower mass neutrinos due to the
production of these momenta during the slow temporal evolution of the
temperature through the quark-hadron transition.

\begin{figure}
\includegraphics[width=3.3truein]{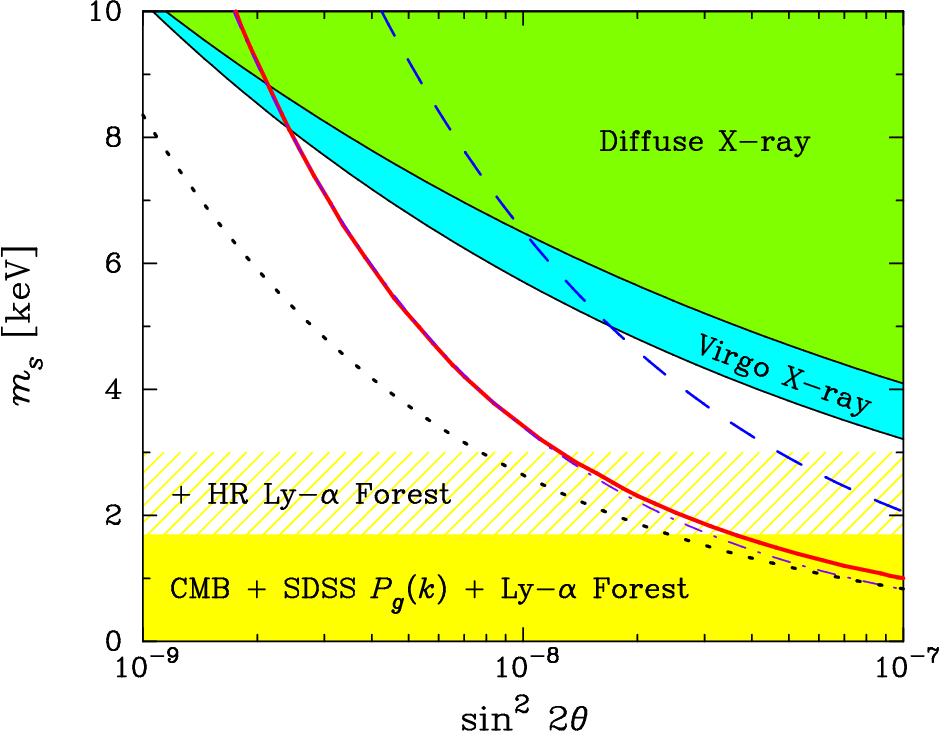}
\caption {\small Contours of predicted critical density $\Omega_{\rm
    DM} =0.26$ from the direct numerical calculation in this work
  (solid red), the fit provided here (dot-dashed purple), the results
  of Ref.~\cite{Abazajian:2001nj} which used a different quark-hadron
  transition and particle population model (dashed blue), and that of
  Ref.~\cite{Dolgov:2000ew} (dotted black) for the inaccurate but
  common choice of $g_*^\prime = 10.75$ (more realistic choices of
  $g_*^\prime$ make the predicted abundance more inaccurate). Also
  shown are the upper flux constraint from X-ray observations of the
  Virgo cluster~\cite{Abazajian:2001vt}, the constraint from the
  diffuse X-ray background~\cite{Boyarsky:2005us}, the lower mass
  constraint from the CMB, SDSS galaxy clustering and Lyman-$\alpha$
  forest, and the possible constraint from also including the
  high-resolution (HR) Lyman-$\alpha$ forest \cite{AbazajianLower05}.}
\label{omega}
\end{figure}

The work on sterile neutrino dark matter production of
Ref.~\cite{Dolgov:2000ew} follows that of Ref.~\cite{Dodelson:1993je}
except for an extension to $m_s > 0.5\rm\ keV$ via the inclusion of an
unknown factor $(g_*^\prime/10.75)$ within the production
``prediction'' relationship, with an effective production statistical
degree of freedom, $g_*^\prime$.  The resulting relation lacked
predictivity due to the power-law dependence on an unknown
$g_*^\prime$. The major effect of the high degrees of freedom at the
production temperatures is dilution of the sterile neutrinos, so that
the production dependence on an increased $g_*^\prime$ is an inverse
relation. The sign and value of the dependency on $g_*^\prime$ in the
production prediction equation of Ref.~\cite{Dolgov:2000ew} was noted
in Ref.~\cite{Viel:2005qj} as a typographical error.  Using the common
but incorrect choice of $g_*^\prime = 10.75$, the production relation
in Ref.~\cite{Dolgov:2000ew} is inaccurate, as shown in
Fig.~\ref{omega}.

In Fig.~\ref{omega}, I show the contours of critical density in
sterile neutrino dark matter density for varying mass and mixing for
an electron neutrino flavor mixed with the sterile neutrino.  I also
show the results of Ref.~\cite{Dolgov:2000ew} for the common choice of
$g_*^\prime = 10.75$, which is inaccurate, but choices of a more
realistic effective production temperature $g_*^\prime$ increases the
discrepancy.  I also show the results of Ref.~\cite{Abazajian:2001nj},
which used $T_{\rm QCD} = 100\rm\ MeV$ and a different model of the
hadron and lepton population distribution based on an older catalogue
of particle masses.  The critical density contour of
Ref.~\cite{Abazajian:2001nj} lies at higher mixing angles because an
increased coupling was required to offset the dilution of the QCD
transition set at $T_{\rm QCD} = 100\rm\ MeV$ , which occurs below the
bulk of production of all sterile neutrino masses $m_s \gtrsim
1\rm\ keV$.  The full numerical results for the predicted dark matter
abundance from this work are fit well by the relation
\begin{eqnarray}
m_s &=& 3.40\ {\rm keV}\ \left(\frac{\sin^2
  2\theta}{10^{-8}}\right)^{-0.615}\ \left(\frac{\Omega_{\rm DM}
  }{0.26}\right)^{0.5}\cr
&&\times\ \left\{ 0.527\ {\rm erfc}\left[ -1.15 \left(\frac{T_{\rm
        QCD}}{170\rm\ MeV}\right)^{2.15}\right]\right\},
\label{prediction}
\end{eqnarray}
for an electron neutrino flavor mixed with the sterile neutrino.  This
expression is valid for $135 < T_{\rm QCD} <
300\rm\ MeV$. Note that the quantity within curled brackets is unity
for $ T_{\rm QCD} = 170\rm\ MeV$.

This new relationship between the critical density fraction, $m_s$ and
$\sin^2 2\theta$ modifies the flux constraint observed from the Virgo
cluster~\cite{Abazajian:2001vt}, so that the inferred upper bound on
the mass of the sterile neutrino dark matter is now
\begin{equation}
m_s < 8.2\rm\ keV.
\end{equation}
This is the result of the fact that the constraint in
Ref.~\cite{Abazajian:2001vt} is not a direct mass constraint but a
flux constraint, which is related to the radiative decay rate through
the mass-mixing angle relation.  Lines of constant flux follow the
fourth power of the mass since the decay rate increases as the fifth
power, but the number density in the field of view decreases
proportionally with the mass.  Using the production relationship,
Eq.~(\ref{prediction}), the diffuse X-ray background limit of
Ref.~\cite{Boyarsky:2005us} is 
\begin{equation}
m_s < 8.89{\rm\ keV} \ \left(\frac{\Omega_{\rm DM}}{0.26}\right)^{0.538},
\label{upperdiffuse}
\end{equation}
for central values of the cosmological parameters, and is shown in
Fig.~\ref{omega}.  The constraints from unresolved X-ray sources
derived by Mapelli \& Ferrara \cite{Mapelli:2005hq} are similar to
Eq.~(\ref{upperdiffuse}), when using the production relation
Eq.~(\ref{prediction})~\cite{Mapelliprivate}.


\section{Perturbation Evolution}
\label{perturbations}
The standard cosmological model of structure formation from adiabatic
Gaussian fluctuations seeded by an inflationary epoch is affected by
perturbation growth in the radiation through matter dominated eras.
The distribution of velocities of the dark matter suppresses
fluctuations below its free streaming scale, which increases with the
mean dark matter velocities and decreases with its mass.  Since
sterile neutrinos are produced non-thermally, their full energy
distribution must be included in an accurate calculation of the
fluctuation spectrum arising from the linear growth epoch.  I use
the approach of the covariant multipole perturbation evolution
equations for massive neutrinos in Ref.~\cite{Lewis:2002nc} and
implemented in the Code for Anisotropies in the Microwave Background
(CAMB)~\cite{Lewis:1999bs}.  The multipole equations depend on the
value of the massive neutrino energy distribution and its momentum
derivative, but I will not reproduce them here.

\begin{figure}
\includegraphics[width=3.3truein]{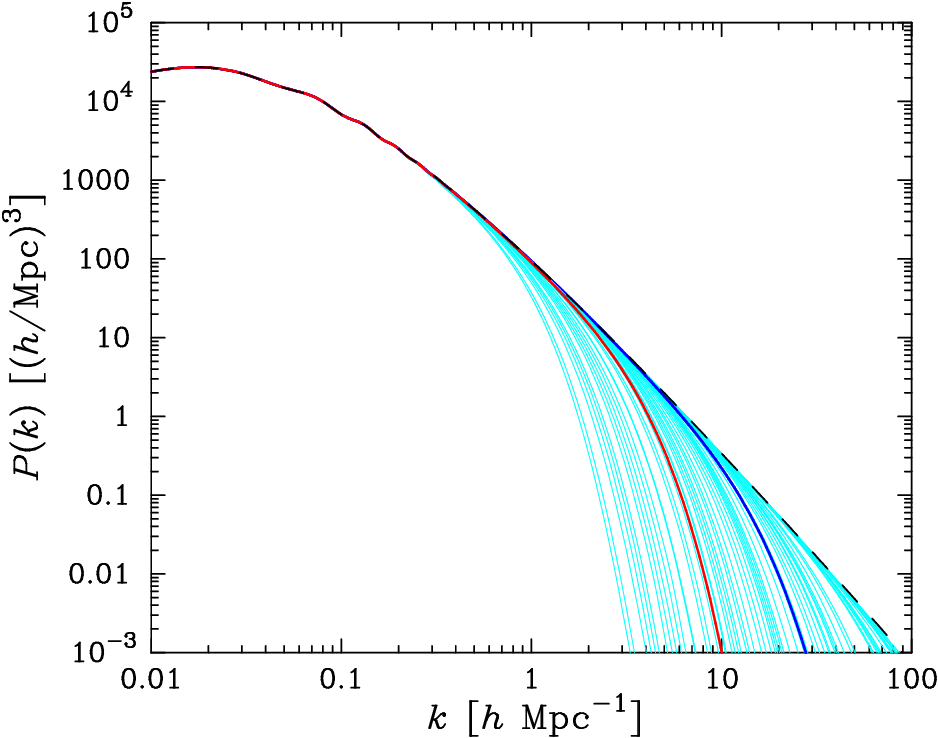}
\caption
{\small Shown are the resulting linear matter power spectra for
  nonthermal sterile neutrinos in the mass range $0.3 < m_s
  < 140{\rm\ keV}$ (gray/cyan).  The thick (red) low-$k$ suppression case
  is for the lower-mass limit inferred from the Lyman-$\alpha$ forest
  ($m_s > 1.7\rm\ keV$) , and the thick (blue) high-$k$ suppression case
  is for the upper-mass limit from X-ray observations of the Virgo
  cluster ($m_s < 8.2\rm\ keV$).  The CDM case is the dashed (black)
  line. Measures of large-scale structure in the linear regime are in the
    region of $0.01 h{\rm\ Mpc}^{-1}<k<0.2 h{\rm\ Mpc}^{-1}$ for
    galaxy surveys, while neutral gas clustering observed in the
    Ly-$\alpha$ forest may extend observations of linear structure to
    $0.1 h{\rm\ Mpc}^{-1}<k<3 h{\rm\ Mpc^{-1}}$.}
\label{matterpower}
\end{figure}

I calculate the growth of perturbations through the radiation and
matter dominated epochs of sterile neutrino dark matter with CAMB.  I
include directly the numerically-calculated momentum-dependent sterile
neutrino distribution functions and their derivatives from the
solution of the quasi-classical Boltzmann Eq.~(\ref{fullboltz}) as
described in the previous section.  The resulting linear matter
power spectra today at redshift zero are shown in
Fig.~\ref{matterpower} for a range of sterile neutrino masses from 0.3
to 140 keV, along with the related CDM case.

\begin{figure}
\includegraphics[width=3.3truein]{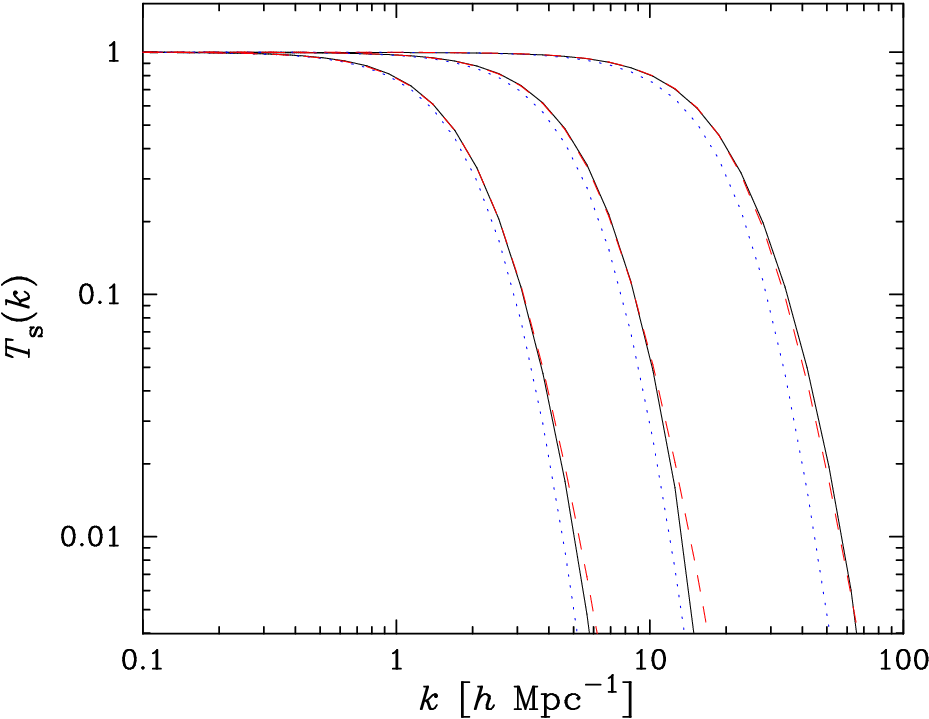}
\caption
{\small Shown here are the relative sterile neutrino transfer function
  $T_s(k)$ to CDM for the same large scale amplitude of perturbations,
  for the cases of $m_s = 0.5, 1.7, 8.2\rm\ keV$ with increasing wave
  number $k$ suppression scale, respectively.  The solid (black) lines
  are from the full numerical calculation, the dashed (red) lines are
  fitting form in Eq.~(\ref{transfer_sterile_fit}), and the dotted
  (blue) lines are the results of Ref.~\cite{Viel:2005qj}.}
\label{transfer_ratio_comparison}
\end{figure}

A useful form of the suppressed perturbation power spectrum $P_{\rm
  sterile}(k)$ relative to the CDM case is a sterile neutrino transfer
function of the form
\begin{equation}
T_s(k) \equiv \sqrt{\frac{P_{\rm sterile}(k)}{P_{\rm CDM}(k)}},
\label{sterile_transfer}
\end{equation}
which can be used to convert any CDM transfer function to that of
sterile neutrino dark matter.  I find a fitting function that
describes the transfer function of the form
\begin{equation}
T_s(k) = \left[1 + \left(\alpha k\right)^\nu\right]^{-\mu}, 
\label{transfer_sterile_fit}
\end{equation}
where 
\begin{equation}
\alpha = a\ \left(\frac{m_s}{1\rm\ keV}\right)^b
\left(\frac{\Omega_{\rm DM}}{0.26}\right)^c
\left(\frac{h}{0.7}\right)^d\ h^{-1}\rm\ Mpc, 
\end{equation}
and $a = 0.189$, $b=-0.858$, $c = -0.136$, $d = 0.692$, $\nu = 2.25$,
and $\mu = 3.08$.  The fitting form is valid for $0.3
\lesssim m_s \lesssim 15\rm\ keV$.  This fitting function is shown
relative to the numerical results in
Fig.~\ref{transfer_ratio_comparison} as well as previous results by
Ref.~\cite{Viel:2005qj}.  Note that all of the features of the
numerical results are not obtained in the fit due to the nonthermal
character of the sterile neutrino distribution, particularly for $m_s
= 1.7 \rm\ keV$ where peak production occurs near the quark-hadron
transition.

The result presented here for the relative sterile neutrino transfer
function is similar, yet significantly different from previous
work~\cite{Hansen:2001zv,Viel:2005qj}, with the difference attributed
the use here of the non-thermal sterile neutrino momentum distribution
due to the physics described in \S\ref{production}.  The results
derived here differ in cosmological parameter dependence of $T_s(k)$
from 2\% to 18\% and in the rapidity of the cutoff $\mu$ at 45\%
relative to that in Refs.~\cite{Hansen:2001zv,Viel:2005qj}.  Using the
transfer function derived here and small scale clustering data sets
including the inferred matter power spectrum from the high-resolution
Lyman-$\alpha$ forest from Viel et al.~\cite{Viel:2004bf},
Ref.~\cite{AbazajianLower05} found lower limits on the mass of the
sterile neutrino dark matter at $1.7\rm\ keV\ (95\% CL)$ from the CMB,
the SDSS 3D $P_g(k)$ of galaxies~\cite{Tegmark:2003uf} plus SDSS
Lyman-$\alpha$ forest~\cite{McDonald:2004xn}, and a lower limit of
$3.0\rm\ keV\ (95\% CL)$ if the inferred matter power spectrum from
the high-resolution Lyman-$\alpha$ forest of Ref.~\cite{Viel:2004bf}
is used, which however has significant systematic uncertainties.

\section{Conclusions}
\label{conclusions}

Potential problems in galaxy and small-scale structure formation
indicate the possibility of a small-scale velocity damping of
perturbations of the type in warm dark matter such as sterile neutrino
dark matter.  Here, I have presented the calculation of the production
and linear perturbation evolution in the early universe of sterile
neutrino dark matter.  Included were the essential effects of the
change in Friedmann dynamics dictated by the quark-hadron transition
and particle population, the alteration of finite temperature
effective mass of active neutrinos due to the presence of thermal
leptons, and heating of the coupled species due to the disappearance
of degrees of freedom in the plasma.  The resulting sterile neutrinos
have a non-trivial momentum distribution that is grossly nonthermal.

Using the resulting energy distributions of sterile neutrinos, I have
calculated the evolution of sterile neutrino density perturbations in
the early universe through the linear regime and provide a fitting
function form for the transfer function describing the suppression of
small scale fluctuations for this warm dark matter candidate.  The
results presented here differ significantly from previous work due to
the inclusion of relevant physical effects. 

The results of the linear perturbation evolution presented here are
necessary as initial conditions for addressing the questions of
structure formation and galaxy formation in the case of a sterile
neutrino dark matter candidate with a non-trivial velocity
distribution.  Following structure formation in this case into the
nonlinear regime will allow a resolution of the question of whether
observations of galactic structure at small scales are indicating a
method of inferring the properties and---ultimately---the identity of
the dark matter.

\acknowledgments

I thank Antony Lewis for assistance with CAMB, Gerard Jungman for help
with rational function fitting forms, Tanmoy Bhattacharya and Rajan
Gupta for useful discussions regarding lattice methods in studies of
the quark-hadron transition, and Salman Habib and Julien Lesgourgues
for very useful comments on the manuscript.  This work was supported
by Los Alamos National Laboratory under DOE contract W-7405-ENG-36.

\bibliography{sndm}

\begin{thebibliography}{64}
\expandafter\ifx\csname natexlab\endcsname\relax\def\natexlab#1{#1}\fi
\expandafter\ifx\csname bibnamefont\endcsname\relax
  \def\bibnamefont#1{#1}\fi
\expandafter\ifx\csname bibfnamefont\endcsname\relax
  \def\bibfnamefont#1{#1}\fi
\expandafter\ifx\csname citenamefont\endcsname\relax
  \def\citenamefont#1{#1}\fi
\expandafter\ifx\csname url\endcsname\relax
  \def\url#1{\texttt{#1}}\fi
\expandafter\ifx\csname urlprefix\endcsname\relax\def\urlprefix{URL }\fi
\providecommand{\bibinfo}[2]{#2}
\providecommand{\eprint}[2][]{\url{#2}}

\bibitem[{\citenamefont{Spergel et~al.}(2003)}]{Spergel:2003cb}
\bibinfo{author}{\bibfnamefont{D.~N.} \bibnamefont{Spergel}}
  \bibnamefont{et~al.} (\bibinfo{collaboration}{WMAP}),
  \bibinfo{journal}{Astrophys. J. Suppl.} \textbf{\bibinfo{volume}{148}},
  \bibinfo{pages}{175} (\bibinfo{year}{2003}), \eprint{astro-ph/0302209}.

\bibitem[{\citenamefont{Tegmark et~al.}(2004{\natexlab{a}})}]{Tegmark:2003ud}
\bibinfo{author}{\bibfnamefont{M.}~\bibnamefont{Tegmark}} \bibnamefont{et~al.}
  (\bibinfo{collaboration}{SDSS}), \bibinfo{journal}{Phys. Rev.}
  \textbf{\bibinfo{volume}{D69}}, \bibinfo{pages}{103501}
  (\bibinfo{year}{2004}{\natexlab{a}}), \eprint{astro-ph/0310723}.

\bibitem[{\citenamefont{Seljak et~al.}(2005)}]{Seljak:2004xh}
\bibinfo{author}{\bibfnamefont{U.}~\bibnamefont{Seljak}} \bibnamefont{et~al.},
  \bibinfo{journal}{Phys. Rev.} \textbf{\bibinfo{volume}{D71}},
  \bibinfo{pages}{103515} (\bibinfo{year}{2005}), \eprint{astro-ph/0407372}.

\bibitem[{\citenamefont{Dodelson and Widrow}(1994)}]{Dodelson:1993je}
\bibinfo{author}{\bibfnamefont{S.}~\bibnamefont{Dodelson}} \bibnamefont{and}
  \bibinfo{author}{\bibfnamefont{L.~M.} \bibnamefont{Widrow}},
  \bibinfo{journal}{Phys. Rev. Lett.} \textbf{\bibinfo{volume}{72}},
  \bibinfo{pages}{17} (\bibinfo{year}{1994}), \eprint{hep-ph/9303287}.

\bibitem[{\citenamefont{Shi and Fuller}(1999)}]{Shi:1998km}
\bibinfo{author}{\bibfnamefont{X.-d.} \bibnamefont{Shi}} \bibnamefont{and}
  \bibinfo{author}{\bibfnamefont{G.~M.} \bibnamefont{Fuller}},
  \bibinfo{journal}{Phys. Rev. Lett.} \textbf{\bibinfo{volume}{82}},
  \bibinfo{pages}{2832} (\bibinfo{year}{1999}), \eprint{astro-ph/9810076}.

\bibitem[{\citenamefont{Hofmann et~al.}(2001)\citenamefont{Hofmann, Schwarz,
  and Stoecker}}]{Hofmann:2001bi}
\bibinfo{author}{\bibfnamefont{S.}~\bibnamefont{Hofmann}},
  \bibinfo{author}{\bibfnamefont{D.~J.} \bibnamefont{Schwarz}},
  \bibnamefont{and} \bibinfo{author}{\bibfnamefont{H.}~\bibnamefont{Stoecker}},
  \bibinfo{journal}{Phys. Rev.} \textbf{\bibinfo{volume}{D64}},
  \bibinfo{pages}{083507} (\bibinfo{year}{2001}), \eprint{astro-ph/0104173}.

\bibitem[{\citenamefont{Jungman et~al.}(1996)\citenamefont{Jungman,
  Kamionkowski, and Griest}}]{Jungman:1995df}
\bibinfo{author}{\bibfnamefont{G.}~\bibnamefont{Jungman}},
  \bibinfo{author}{\bibfnamefont{M.}~\bibnamefont{Kamionkowski}},
  \bibnamefont{and} \bibinfo{author}{\bibfnamefont{K.}~\bibnamefont{Griest}},
  \bibinfo{journal}{Phys. Rept.} \textbf{\bibinfo{volume}{267}},
  \bibinfo{pages}{195} (\bibinfo{year}{1996}), \eprint{hep-ph/9506380}.

\bibitem[{\citenamefont{Kauffmann et~al.}(1993)\citenamefont{Kauffmann, White,
  and Guiderdoni}}]{Kauffmann:1993gv}
\bibinfo{author}{\bibfnamefont{G.}~\bibnamefont{Kauffmann}},
  \bibinfo{author}{\bibfnamefont{S.~D.~M.} \bibnamefont{White}},
  \bibnamefont{and}
  \bibinfo{author}{\bibfnamefont{B.}~\bibnamefont{Guiderdoni}},
  \bibinfo{journal}{Mon. Not. Roy. Astron. Soc.}
  \textbf{\bibinfo{volume}{264}}, \bibinfo{pages}{201} (\bibinfo{year}{1993}).

\bibitem[{\citenamefont{Klypin et~al.}(1999)\citenamefont{Klypin, Kravtsov,
  Valenzuela, and Prada}}]{Klypin:1999uc}
\bibinfo{author}{\bibfnamefont{A.~A.} \bibnamefont{Klypin}},
  \bibinfo{author}{\bibfnamefont{A.~V.} \bibnamefont{Kravtsov}},
  \bibinfo{author}{\bibfnamefont{O.}~\bibnamefont{Valenzuela}},
  \bibnamefont{and} \bibinfo{author}{\bibfnamefont{F.}~\bibnamefont{Prada}},
  \bibinfo{journal}{Astrophys. J.} \textbf{\bibinfo{volume}{522}},
  \bibinfo{pages}{82} (\bibinfo{year}{1999}), \eprint{astro-ph/9901240}.

\bibitem[{\citenamefont{Moore et~al.}(1999)}]{Moore:1999wf}
\bibinfo{author}{\bibfnamefont{B.}~\bibnamefont{Moore}} \bibnamefont{et~al.},
  \bibinfo{journal}{Astrophys. J.} \textbf{\bibinfo{volume}{524}},
  \bibinfo{pages}{L19} (\bibinfo{year}{1999}), \eprint{astro-ph/9907411}.

\bibitem[{\citenamefont{Willman et~al.}(2004)\citenamefont{Willman, Governato,
  Wadsley, and Quinn}}]{Willman:2004xc}
\bibinfo{author}{\bibfnamefont{B.}~\bibnamefont{Willman}},
  \bibinfo{author}{\bibfnamefont{F.}~\bibnamefont{Governato}},
  \bibinfo{author}{\bibfnamefont{J.}~\bibnamefont{Wadsley}}, \bibnamefont{and}
  \bibinfo{author}{\bibfnamefont{T.}~\bibnamefont{Quinn}},
  \bibinfo{journal}{Mon. Not. Roy. Astron. Soc.}
  \textbf{\bibinfo{volume}{353}}, \bibinfo{pages}{639} (\bibinfo{year}{2004}),
  \eprint{astro-ph/0403001}.

\bibitem[{\citenamefont{Bode et~al.}(2001)\citenamefont{Bode, Ostriker, and
  Turok}}]{Bode:2000gq}
\bibinfo{author}{\bibfnamefont{P.}~\bibnamefont{Bode}},
  \bibinfo{author}{\bibfnamefont{J.~P.} \bibnamefont{Ostriker}},
  \bibnamefont{and} \bibinfo{author}{\bibfnamefont{N.}~\bibnamefont{Turok}},
  \bibinfo{journal}{Astrophys. J.} \textbf{\bibinfo{volume}{556}},
  \bibinfo{pages}{93} (\bibinfo{year}{2001}), \eprint{astro-ph/0010389}.

\bibitem[{\citenamefont{Dekel and Silk}(1986)}]{Dekel:1986gu}
\bibinfo{author}{\bibfnamefont{A.}~\bibnamefont{Dekel}} \bibnamefont{and}
  \bibinfo{author}{\bibfnamefont{J.}~\bibnamefont{Silk}},
  \bibinfo{journal}{Astrophys. J.} \textbf{\bibinfo{volume}{303}},
  \bibinfo{pages}{39} (\bibinfo{year}{1986}).

\bibitem[{\citenamefont{{Thoul} and {Weinberg}}(1996)}]{Thoul}
\bibinfo{author}{\bibfnamefont{A.~A.} \bibnamefont{{Thoul}}} \bibnamefont{and}
  \bibinfo{author}{\bibfnamefont{D.~H.} \bibnamefont{{Weinberg}}},
  \bibinfo{journal}{Astrophys. J.} \textbf{\bibinfo{volume}{465}},
  \bibinfo{pages}{608} (\bibinfo{year}{1996}), \eprint{astro-ph/9510154}.

\bibitem[{\citenamefont{Bullock et~al.}(2001)\citenamefont{Bullock, Kravtsov,
  and Weinberg}}]{Bullock:2000qf}
\bibinfo{author}{\bibfnamefont{J.~S.} \bibnamefont{Bullock}},
  \bibinfo{author}{\bibfnamefont{A.~V.} \bibnamefont{Kravtsov}},
  \bibnamefont{and} \bibinfo{author}{\bibfnamefont{D.~H.}
  \bibnamefont{Weinberg}}, \bibinfo{journal}{Astrophys. J.}
  \textbf{\bibinfo{volume}{548}}, \bibinfo{pages}{33} (\bibinfo{year}{2001}),
  \eprint{astro-ph/0007295}.

\bibitem[{\citenamefont{Bullock et~al.}(2000)\citenamefont{Bullock, Kravtsov,
  and Weinberg}}]{Bullock:2000wn}
\bibinfo{author}{\bibfnamefont{J.~S.} \bibnamefont{Bullock}},
  \bibinfo{author}{\bibfnamefont{A.~V.} \bibnamefont{Kravtsov}},
  \bibnamefont{and} \bibinfo{author}{\bibfnamefont{D.~H.}
  \bibnamefont{Weinberg}}, \bibinfo{journal}{Astrophys. J.}
  \textbf{\bibinfo{volume}{539}}, \bibinfo{pages}{517} (\bibinfo{year}{2000}),
  \eprint{astro-ph/0002214}.

\bibitem[{\citenamefont{Benson et~al.}(2002)\citenamefont{Benson, Lacey, Baugh,
  Cole, and Frenk}}]{Benson:2001au}
\bibinfo{author}{\bibfnamefont{A.~J.} \bibnamefont{Benson}},
  \bibinfo{author}{\bibfnamefont{C.~G.} \bibnamefont{Lacey}},
  \bibinfo{author}{\bibfnamefont{C.~M.} \bibnamefont{Baugh}},
  \bibinfo{author}{\bibfnamefont{S.}~\bibnamefont{Cole}}, \bibnamefont{and}
  \bibinfo{author}{\bibfnamefont{C.~S.} \bibnamefont{Frenk}},
  \bibinfo{journal}{Mon. Not. Roy. Astron. Soc.}
  \textbf{\bibinfo{volume}{333}}, \bibinfo{pages}{156} (\bibinfo{year}{2002}),
  \eprint{astro-ph/0108217}.

\bibitem[{\citenamefont{Peebles}(2001)}]{Peebles:2001nv}
\bibinfo{author}{\bibfnamefont{P.~J.~E.} \bibnamefont{Peebles}}
  (\bibinfo{year}{2001}), \eprint{astro-ph/0101127}.

\bibitem[{\citenamefont{Dalcanton and Hogan}(2001)}]{Dalcanton:2000hn}
\bibinfo{author}{\bibfnamefont{J.~J.} \bibnamefont{Dalcanton}}
  \bibnamefont{and} \bibinfo{author}{\bibfnamefont{C.~J.} \bibnamefont{Hogan}},
  \bibinfo{journal}{Astrophys. J.} \textbf{\bibinfo{volume}{561}},
  \bibinfo{pages}{35} (\bibinfo{year}{2001}), \eprint{astro-ph/0004381}.

\bibitem[{\citenamefont{van~den Bosch and Swaters}(2001)}]{vandenBosch:2000rz}
\bibinfo{author}{\bibfnamefont{F.~C.} \bibnamefont{van~den Bosch}}
  \bibnamefont{and} \bibinfo{author}{\bibfnamefont{R.~A.}
  \bibnamefont{Swaters}}, \bibinfo{journal}{Mon. Not. Roy. Astron. Soc.}
  \textbf{\bibinfo{volume}{325}}, \bibinfo{pages}{1017} (\bibinfo{year}{2001}),
  \eprint{astro-ph/0006048}.

\bibitem[{\citenamefont{Zentner and Bullock}(2002)}]{Zentner:2002xt}
\bibinfo{author}{\bibfnamefont{A.~R.} \bibnamefont{Zentner}} \bibnamefont{and}
  \bibinfo{author}{\bibfnamefont{J.~S.} \bibnamefont{Bullock}},
  \bibinfo{journal}{Phys. Rev.} \textbf{\bibinfo{volume}{D66}},
  \bibinfo{pages}{043003} (\bibinfo{year}{2002}), \eprint{astro-ph/0205216}.

\bibitem[{\citenamefont{Abazajian et~al.}(2005)\citenamefont{Abazajian,
  Koushiappas, and Zentner}}]{Abazajian:2005kz}
\bibinfo{author}{\bibfnamefont{K.}~\bibnamefont{Abazajian}},
  \bibinfo{author}{\bibfnamefont{S.~M.} \bibnamefont{Koushiappas}},
  \bibnamefont{and} \bibinfo{author}{\bibfnamefont{A.~R.}
  \bibnamefont{Zentner}}, \bibinfo{journal}{in preparation}
  (\bibinfo{year}{2005}).

\bibitem[{\citenamefont{Dolgov and Sommer-Larsen}(2001)}]{Dolgov:2001nq}
\bibinfo{author}{\bibfnamefont{A.~D.} \bibnamefont{Dolgov}} \bibnamefont{and}
  \bibinfo{author}{\bibfnamefont{J.}~\bibnamefont{Sommer-Larsen}},
  \bibinfo{journal}{Astrophys. J.} \textbf{\bibinfo{volume}{551}},
  \bibinfo{pages}{608} (\bibinfo{year}{2001}).

\bibitem[{\citenamefont{Governato et~al.}(2004)}]{Governato:2002cv}
\bibinfo{author}{\bibfnamefont{F.}~\bibnamefont{Governato}}
  \bibnamefont{et~al.}, \bibinfo{journal}{Astrophys. J.}
  \textbf{\bibinfo{volume}{607}}, \bibinfo{pages}{688} (\bibinfo{year}{2004}),
  \eprint{astro-ph/0207044}.

\bibitem[{\citenamefont{{Kormendy} and {Fisher}}(2005)}]{Kormendy2005}
\bibinfo{author}{\bibfnamefont{J.}~\bibnamefont{{Kormendy}}} \bibnamefont{and}
  \bibinfo{author}{\bibfnamefont{D.~B.} \bibnamefont{{Fisher}}},
  \bibinfo{journal}{Revista Mexicana de Astronomia y Astrofisica (Conference
  Series)} \textbf{\bibinfo{volume}{25}}, \bibinfo{pages}{101}
  (\bibinfo{year}{2005}), \eprint{astro-ph/0507525}.

\bibitem[{\citenamefont{Brahmachari and Mohapatra}(1998)}]{Brahmachari:1998kt}
\bibinfo{author}{\bibfnamefont{B.}~\bibnamefont{Brahmachari}} \bibnamefont{and}
  \bibinfo{author}{\bibfnamefont{R.~N.} \bibnamefont{Mohapatra}},
  \bibinfo{journal}{Phys. Lett.} \textbf{\bibinfo{volume}{B437}},
  \bibinfo{pages}{100} (\bibinfo{year}{1998}), \eprint{hep-ph/9805429}.

\bibitem[{\citenamefont{Langacker}(1998)}]{Langacker:1998ut}
\bibinfo{author}{\bibfnamefont{P.}~\bibnamefont{Langacker}},
  \bibinfo{journal}{Phys. Rev.} \textbf{\bibinfo{volume}{D58}},
  \bibinfo{pages}{093017} (\bibinfo{year}{1998}), \eprint{hep-ph/9805281}.

\bibitem[{\citenamefont{Dvali and Smirnov}(1999)}]{Dvali:1999cn}
\bibinfo{author}{\bibfnamefont{G.~R.} \bibnamefont{Dvali}} \bibnamefont{and}
  \bibinfo{author}{\bibfnamefont{A.~Y.} \bibnamefont{Smirnov}},
  \bibinfo{journal}{Nucl. Phys.} \textbf{\bibinfo{volume}{B563}},
  \bibinfo{pages}{63} (\bibinfo{year}{1999}), \eprint{hep-ph/9904211}.

\bibitem[{\citenamefont{Abazajian et~al.}(2003)\citenamefont{Abazajian, Fuller,
  and Patel}}]{Abazajian:2000hw}
\bibinfo{author}{\bibfnamefont{K.}~\bibnamefont{Abazajian}},
  \bibinfo{author}{\bibfnamefont{G.~M.} \bibnamefont{Fuller}},
  \bibnamefont{and} \bibinfo{author}{\bibfnamefont{M.}~\bibnamefont{Patel}},
  \bibinfo{journal}{Phys. Rev. Lett.} \textbf{\bibinfo{volume}{90}},
  \bibinfo{pages}{061301} (\bibinfo{year}{2003}), \eprint{hep-ph/0011048}.

\bibitem[{\citenamefont{Asaka et~al.}(2005)\citenamefont{Asaka, Blanchet, and
  Shaposhnikov}}]{Asaka:2005an}
\bibinfo{author}{\bibfnamefont{T.}~\bibnamefont{Asaka}},
  \bibinfo{author}{\bibfnamefont{S.}~\bibnamefont{Blanchet}}, \bibnamefont{and}
  \bibinfo{author}{\bibfnamefont{M.}~\bibnamefont{Shaposhnikov}}
  (\bibinfo{year}{2005}), \eprint{hep-ph/0503065}.

\bibitem[{\citenamefont{Athanassopoulos et~al.}(1998)}]{Athanassopoulos:1997pv}
\bibinfo{author}{\bibfnamefont{C.}~\bibnamefont{Athanassopoulos}}
  \bibnamefont{et~al.} (\bibinfo{collaboration}{LSND}), \bibinfo{journal}{Phys.
  Rev. Lett.} \textbf{\bibinfo{volume}{81}}, \bibinfo{pages}{1774}
  (\bibinfo{year}{1998}), \eprint{nucl-ex/9709006}.

\bibitem[{\citenamefont{Sorel et~al.}(2004)\citenamefont{Sorel, Conrad, and
  Shaevitz}}]{Sorel:2003hf}
\bibinfo{author}{\bibfnamefont{M.}~\bibnamefont{Sorel}},
  \bibinfo{author}{\bibfnamefont{J.~M.} \bibnamefont{Conrad}},
  \bibnamefont{and} \bibinfo{author}{\bibfnamefont{M.}~\bibnamefont{Shaevitz}},
  \bibinfo{journal}{Phys. Rev.} \textbf{\bibinfo{volume}{D70}},
  \bibinfo{pages}{073004} (\bibinfo{year}{2004}), \eprint{hep-ph/0305255}.

\bibitem[{\citenamefont{Kusenko and Segre}(1999)}]{Kusenko:1998bk}
\bibinfo{author}{\bibfnamefont{A.}~\bibnamefont{Kusenko}} \bibnamefont{and}
  \bibinfo{author}{\bibfnamefont{G.}~\bibnamefont{Segre}},
  \bibinfo{journal}{Phys. Rev.} \textbf{\bibinfo{volume}{D59}},
  \bibinfo{pages}{061302} (\bibinfo{year}{1999}), \eprint{astro-ph/9811144}.

\bibitem[{\citenamefont{Fuller et~al.}(2003)\citenamefont{Fuller, Kusenko,
  Mocioiu, and Pascoli}}]{Fuller:2003gy}
\bibinfo{author}{\bibfnamefont{G.~M.} \bibnamefont{Fuller}},
  \bibinfo{author}{\bibfnamefont{A.}~\bibnamefont{Kusenko}},
  \bibinfo{author}{\bibfnamefont{I.}~\bibnamefont{Mocioiu}}, \bibnamefont{and}
  \bibinfo{author}{\bibfnamefont{S.}~\bibnamefont{Pascoli}},
  \bibinfo{journal}{Phys. Rev.} \textbf{\bibinfo{volume}{D68}},
  \bibinfo{pages}{103002} (\bibinfo{year}{2003}), \eprint{astro-ph/0307267}.

\bibitem[{\citenamefont{Kusenko}(2004)}]{Kusenko:2004mm}
\bibinfo{author}{\bibfnamefont{A.}~\bibnamefont{Kusenko}},
  \bibinfo{journal}{Int. J. Mod. Phys.} \textbf{\bibinfo{volume}{D13}},
  \bibinfo{pages}{2065} (\bibinfo{year}{2004}), \eprint{astro-ph/0409521}.

\bibitem[{\citenamefont{Scheck et~al.}(2004)\citenamefont{Scheck, Plewa, Janka,
  Kifonidis, and Mueller}}]{Scheck:2003rw}
\bibinfo{author}{\bibfnamefont{L.}~\bibnamefont{Scheck}},
  \bibinfo{author}{\bibfnamefont{T.}~\bibnamefont{Plewa}},
  \bibinfo{author}{\bibfnamefont{H.-T.} \bibnamefont{Janka}},
  \bibinfo{author}{\bibfnamefont{K.}~\bibnamefont{Kifonidis}},
  \bibnamefont{and} \bibinfo{author}{\bibfnamefont{E.}~\bibnamefont{Mueller}},
  \bibinfo{journal}{Phys. Rev. Lett.} \textbf{\bibinfo{volume}{92}},
  \bibinfo{pages}{011103} (\bibinfo{year}{2004}), \eprint{astro-ph/0307352}.

\bibitem[{\citenamefont{Fryer and Kusenko}(2005)}]{Fryer:2005sz}
\bibinfo{author}{\bibfnamefont{C.~L.} \bibnamefont{Fryer}} \bibnamefont{and}
  \bibinfo{author}{\bibfnamefont{A.}~\bibnamefont{Kusenko}}
  (\bibinfo{year}{2005}), \eprint{astro-ph/0512033}.

\bibitem[{\citenamefont{{Barkana} et~al.}(2001)\citenamefont{{Barkana},
  {Haiman}, and {Ostriker}}}]{Barkana2001}
\bibinfo{author}{\bibfnamefont{R.}~\bibnamefont{{Barkana}}},
  \bibinfo{author}{\bibfnamefont{Z.}~\bibnamefont{{Haiman}}}, \bibnamefont{and}
  \bibinfo{author}{\bibfnamefont{J.~P.} \bibnamefont{{Ostriker}}},
  \bibinfo{journal}{Astrophys. J.} \textbf{\bibinfo{volume}{558}},
  \bibinfo{pages}{482} (\bibinfo{year}{2001}), \eprint{astro-ph/0102304}.

\bibitem[{\citenamefont{Yoshida et~al.}(2003)\citenamefont{Yoshida, Sokasian,
  Hernquist, and Springel}}]{Yoshida:2003rm}
\bibinfo{author}{\bibfnamefont{N.}~\bibnamefont{Yoshida}},
  \bibinfo{author}{\bibfnamefont{A.}~\bibnamefont{Sokasian}},
  \bibinfo{author}{\bibfnamefont{L.}~\bibnamefont{Hernquist}},
  \bibnamefont{and} \bibinfo{author}{\bibfnamefont{V.}~\bibnamefont{Springel}},
  \bibinfo{journal}{Astrophys. J.} \textbf{\bibinfo{volume}{591}},
  \bibinfo{pages}{L1} (\bibinfo{year}{2003}), \eprint{astro-ph/0303622}.

\bibitem[{\citenamefont{Biermann and Kusenko}(2006)}]{Biermann:2006bu}
\bibinfo{author}{\bibfnamefont{P.~L.} \bibnamefont{Biermann}} \bibnamefont{and}
  \bibinfo{author}{\bibfnamefont{A.}~\bibnamefont{Kusenko}},
  \bibinfo{journal}{[Phys.\ Rev.\ Lett.\ (to be published)]}
  (\bibinfo{year}{2006}), \eprint{astro-ph/0601004}.

\bibitem[{\citenamefont{Narayanan et~al.}(2000)\citenamefont{Narayanan,
  Spergel, Dave, and Ma}}]{Narayanan:2000tp}
\bibinfo{author}{\bibfnamefont{V.~K.} \bibnamefont{Narayanan}},
  \bibinfo{author}{\bibfnamefont{D.~N.} \bibnamefont{Spergel}},
  \bibinfo{author}{\bibfnamefont{R.}~\bibnamefont{Dave}}, \bibnamefont{and}
  \bibinfo{author}{\bibfnamefont{C.-P.} \bibnamefont{Ma}},
  \bibinfo{journal}{Astrophys. J.} \textbf{\bibinfo{volume}{543}},
  \bibinfo{pages}{L103} (\bibinfo{year}{2000}), \eprint{astro-ph/0005095}.

\bibitem[{\citenamefont{Abazajian}(2006)}]{AbazajianLower05}
\bibinfo{author}{\bibfnamefont{K.}~\bibnamefont{Abazajian}},
  \bibinfo{journal}{[Phys.\ Rev.\ D (to be published)]}
  (\bibinfo{year}{2006}), \eprint{astro-ph/0512631}.

\bibitem[{\citenamefont{Hansen et~al.}(2002)\citenamefont{Hansen, Lesgourgues,
  Pastor, and Silk}}]{Hansen:2001zv}
\bibinfo{author}{\bibfnamefont{S.~H.} \bibnamefont{Hansen}},
  \bibinfo{author}{\bibfnamefont{J.}~\bibnamefont{Lesgourgues}},
  \bibinfo{author}{\bibfnamefont{S.}~\bibnamefont{Pastor}}, \bibnamefont{and}
  \bibinfo{author}{\bibfnamefont{J.}~\bibnamefont{Silk}},
  \bibinfo{journal}{Mon. Not. Roy. Astron. Soc.}
  \textbf{\bibinfo{volume}{333}}, \bibinfo{pages}{544} (\bibinfo{year}{2002}),
  \eprint{astro-ph/0106108}.

\bibitem[{\citenamefont{Viel et~al.}(2005)\citenamefont{Viel, Lesgourgues,
  Haehnelt, Matarrese, and Riotto}}]{Viel:2005qj}
\bibinfo{author}{\bibfnamefont{M.}~\bibnamefont{Viel}},
  \bibinfo{author}{\bibfnamefont{J.}~\bibnamefont{Lesgourgues}},
  \bibinfo{author}{\bibfnamefont{M.~G.} \bibnamefont{Haehnelt}},
  \bibinfo{author}{\bibfnamefont{S.}~\bibnamefont{Matarrese}},
  \bibnamefont{and} \bibinfo{author}{\bibfnamefont{A.}~\bibnamefont{Riotto}},
  \bibinfo{journal}{Phys. Rev.} \textbf{\bibinfo{volume}{D71}},
  \bibinfo{pages}{063534} (\bibinfo{year}{2005}), \eprint{astro-ph/0501562}.

\bibitem[{\citenamefont{McKellar and Thomson}(1994)}]{McKellar:1992ja}
\bibinfo{author}{\bibfnamefont{B.~H.~J.} \bibnamefont{McKellar}}
  \bibnamefont{and} \bibinfo{author}{\bibfnamefont{M.~J.}
  \bibnamefont{Thomson}}, \bibinfo{journal}{Phys. Rev.}
  \textbf{\bibinfo{volume}{D49}}, \bibinfo{pages}{2710} (\bibinfo{year}{1994}).

\bibitem[{\citenamefont{Abazajian
  et~al.}(2001{\natexlab{a}})\citenamefont{Abazajian, Fuller, and
  Patel}}]{Abazajian:2001nj}
\bibinfo{author}{\bibfnamefont{K.}~\bibnamefont{Abazajian}},
  \bibinfo{author}{\bibfnamefont{G.~M.} \bibnamefont{Fuller}},
  \bibnamefont{and} \bibinfo{author}{\bibfnamefont{M.}~\bibnamefont{Patel}},
  \bibinfo{journal}{Phys. Rev.} \textbf{\bibinfo{volume}{D64}},
  \bibinfo{pages}{023501} (\bibinfo{year}{2001}{\natexlab{a}}),
  \eprint{astro-ph/0101524}.

\bibitem[{\citenamefont{Notzold and Raffelt}(1988)}]{Notzold:1987ik}
\bibinfo{author}{\bibfnamefont{D.}~\bibnamefont{Notzold}} \bibnamefont{and}
  \bibinfo{author}{\bibfnamefont{G.}~\bibnamefont{Raffelt}},
  \bibinfo{journal}{Nucl. Phys.} \textbf{\bibinfo{volume}{B307}},
  \bibinfo{pages}{924} (\bibinfo{year}{1988}).

\bibitem[{\citenamefont{Dolgov et~al.}(2002)}]{Dolgov:2002ab}
\bibinfo{author}{\bibfnamefont{A.~D.} \bibnamefont{Dolgov}}
  \bibnamefont{et~al.}, \bibinfo{journal}{Nucl. Phys.}
  \textbf{\bibinfo{volume}{B632}}, \bibinfo{pages}{363} (\bibinfo{year}{2002}),
  \eprint{hep-ph/0201287}.

\bibitem[{\citenamefont{Abazajian et~al.}(2002)\citenamefont{Abazajian, Beacom,
  and Bell}}]{Abazajian:2002qx}
\bibinfo{author}{\bibfnamefont{K.~N.} \bibnamefont{Abazajian}},
  \bibinfo{author}{\bibfnamefont{J.~F.} \bibnamefont{Beacom}},
  \bibnamefont{and} \bibinfo{author}{\bibfnamefont{N.~F.} \bibnamefont{Bell}},
  \bibinfo{journal}{Phys. Rev.} \textbf{\bibinfo{volume}{D66}},
  \bibinfo{pages}{013008} (\bibinfo{year}{2002}), \eprint{astro-ph/0203442}.

\bibitem[{\citenamefont{Wong}(2002)}]{Wong:2002fa}
\bibinfo{author}{\bibfnamefont{Y.~Y.~Y.} \bibnamefont{Wong}},
  \bibinfo{journal}{Phys. Rev.} \textbf{\bibinfo{volume}{D66}},
  \bibinfo{pages}{025015} (\bibinfo{year}{2002}), \eprint{hep-ph/0203180}.

\bibitem[{\citenamefont{Eidelman et~al.}(2004)}]{PDBook}
\bibinfo{author}{\bibfnamefont{S.}~\bibnamefont{Eidelman}}
  \bibnamefont{et~al.}, \bibinfo{journal}{{Physics Letters B}}
  \textbf{\bibinfo{volume}{592}}, \bibinfo{pages}{1+} (\bibinfo{year}{2004}),
  \urlprefix\url{http://pdg.lbl.gov}.

\bibitem[{\citenamefont{Abazajian and Fuller}(2002)}]{Abazajian:2002yz}
\bibinfo{author}{\bibfnamefont{K.~N.} \bibnamefont{Abazajian}}
  \bibnamefont{and} \bibinfo{author}{\bibfnamefont{G.~M.}
  \bibnamefont{Fuller}}, \bibinfo{journal}{Phys. Rev.}
  \textbf{\bibinfo{volume}{D66}}, \bibinfo{pages}{023526}
  (\bibinfo{year}{2002}), \eprint{astro-ph/0204293}.

\bibitem[{\citenamefont{Karsch et~al.}(2001)\citenamefont{Karsch, Laermann, and
  Peikert}}]{Karsch:2000kv}
\bibinfo{author}{\bibfnamefont{F.}~\bibnamefont{Karsch}},
  \bibinfo{author}{\bibfnamefont{E.}~\bibnamefont{Laermann}}, \bibnamefont{and}
  \bibinfo{author}{\bibfnamefont{A.}~\bibnamefont{Peikert}},
  \bibinfo{journal}{Nucl. Phys.} \textbf{\bibinfo{volume}{B605}},
  \bibinfo{pages}{579} (\bibinfo{year}{2001}), \eprint{hep-lat/0012023}.

\bibitem[{\citenamefont{Bernard et~al.}(2005)}]{Bernard:2004je}
\bibinfo{author}{\bibfnamefont{C.}~\bibnamefont{Bernard}} \bibnamefont{et~al.}
  (\bibinfo{collaboration}{MILC}), \bibinfo{journal}{Phys. Rev.}
  \textbf{\bibinfo{volume}{D71}}, \bibinfo{pages}{034504}
  (\bibinfo{year}{2005}), \eprint{hep-lat/0405029}.

\bibitem[{\citenamefont{Dolgov and Hansen}(2002)}]{Dolgov:2000ew}
\bibinfo{author}{\bibfnamefont{A.~D.} \bibnamefont{Dolgov}} \bibnamefont{and}
  \bibinfo{author}{\bibfnamefont{S.~H.} \bibnamefont{Hansen}},
  \bibinfo{journal}{Astropart. Phys.} \textbf{\bibinfo{volume}{16}},
  \bibinfo{pages}{339} (\bibinfo{year}{2002}), \eprint{hep-ph/0009083}.

\bibitem[{\citenamefont{Abazajian
  et~al.}(2001{\natexlab{b}})\citenamefont{Abazajian, Fuller, and
  Tucker}}]{Abazajian:2001vt}
\bibinfo{author}{\bibfnamefont{K.}~\bibnamefont{Abazajian}},
  \bibinfo{author}{\bibfnamefont{G.~M.} \bibnamefont{Fuller}},
  \bibnamefont{and} \bibinfo{author}{\bibfnamefont{W.~H.}
  \bibnamefont{Tucker}}, \bibinfo{journal}{Astrophys. J.}
  \textbf{\bibinfo{volume}{562}}, \bibinfo{pages}{593}
  (\bibinfo{year}{2001}{\natexlab{b}}), \eprint{astro-ph/0106002}.

\bibitem[{\citenamefont{Boyarsky et~al.}(2005)\citenamefont{Boyarsky, Neronov,
  Ruchayskiy, and Shaposhnikov}}]{Boyarsky:2005us}
\bibinfo{author}{\bibfnamefont{A.}~\bibnamefont{Boyarsky}},
  \bibinfo{author}{\bibfnamefont{A.}~\bibnamefont{Neronov}},
  \bibinfo{author}{\bibfnamefont{O.}~\bibnamefont{Ruchayskiy}},
  \bibnamefont{and}
  \bibinfo{author}{\bibfnamefont{M.}~\bibnamefont{Shaposhnikov}}
  (\bibinfo{year}{2005}), \eprint{astro-ph/0512509}.

\bibitem[{\citenamefont{Mapelli and Ferrara}(2005)}]{Mapelli:2005hq}
\bibinfo{author}{\bibfnamefont{M.}~\bibnamefont{Mapelli}} \bibnamefont{and}
  \bibinfo{author}{\bibfnamefont{A.}~\bibnamefont{Ferrara}}
  (\bibinfo{year}{2005}), \eprint{astro-ph/0508413}.

\bibitem[{\citenamefont{Mapelli}(2005)}]{Mapelliprivate}
\bibinfo{author}{\bibfnamefont{M.}~\bibnamefont{Mapelli}}
  (\bibinfo{year}{2005}), \eprint{private communication}.

\bibitem[{\citenamefont{Lewis and Challinor}(2002)}]{Lewis:2002nc}
\bibinfo{author}{\bibfnamefont{A.}~\bibnamefont{Lewis}} \bibnamefont{and}
  \bibinfo{author}{\bibfnamefont{A.}~\bibnamefont{Challinor}},
  \bibinfo{journal}{Phys. Rev.} \textbf{\bibinfo{volume}{D66}},
  \bibinfo{pages}{023531} (\bibinfo{year}{2002}), \eprint{astro-ph/0203507}.

\bibitem[{\citenamefont{Lewis et~al.}(2000)\citenamefont{Lewis, Challinor, and
  Lasenby}}]{Lewis:1999bs}
\bibinfo{author}{\bibfnamefont{A.}~\bibnamefont{Lewis}},
  \bibinfo{author}{\bibfnamefont{A.}~\bibnamefont{Challinor}},
  \bibnamefont{and} \bibinfo{author}{\bibfnamefont{A.}~\bibnamefont{Lasenby}},
  \bibinfo{journal}{Astrophys. J.} \textbf{\bibinfo{volume}{538}},
  \bibinfo{pages}{473} (\bibinfo{year}{2000}), \eprint{astro-ph/9911177},
  \urlprefix\url{http://camb.info}.

\bibitem[{\citenamefont{Viel et~al.}(2004)\citenamefont{Viel, Haehnelt, and
  Springel}}]{Viel:2004bf}
\bibinfo{author}{\bibfnamefont{M.}~\bibnamefont{Viel}},
  \bibinfo{author}{\bibfnamefont{M.~G.} \bibnamefont{Haehnelt}},
  \bibnamefont{and} \bibinfo{author}{\bibfnamefont{V.}~\bibnamefont{Springel}},
  \bibinfo{journal}{Mon. Not. Roy. Astron. Soc.}
  \textbf{\bibinfo{volume}{354}}, \bibinfo{pages}{684} (\bibinfo{year}{2004}),
  \eprint{astro-ph/0404600}.

\bibitem[{\citenamefont{Tegmark et~al.}(2004{\natexlab{b}})}]{Tegmark:2003uf}
\bibinfo{author}{\bibfnamefont{M.}~\bibnamefont{Tegmark}} \bibnamefont{et~al.}
  (\bibinfo{collaboration}{SDSS}), \bibinfo{journal}{Astrophys. J.}
  \textbf{\bibinfo{volume}{606}}, \bibinfo{pages}{702}
  (\bibinfo{year}{2004}{\natexlab{b}}), \eprint{astro-ph/0310725}.

\bibitem[{\citenamefont{{McDonald} et~al.}(2005)\citenamefont{{McDonald},
  {Seljak}, {Cen}, {Shih}, {Weinberg}, {Burles}, {Schneider}, {Schlegel},
  {Bahcall}, {Briggs} et~al.}}]{McDonald:2004xn}
\bibinfo{author}{\bibfnamefont{P.}~\bibnamefont{{McDonald}}},
  \bibinfo{author}{\bibfnamefont{U.}~\bibnamefont{{Seljak}}},
  \bibinfo{author}{\bibfnamefont{R.}~\bibnamefont{{Cen}}},
  \bibinfo{author}{\bibfnamefont{D.}~\bibnamefont{{Shih}}},
  \bibinfo{author}{\bibfnamefont{D.~H.} \bibnamefont{{Weinberg}}},
  \bibinfo{author}{\bibfnamefont{S.}~\bibnamefont{{Burles}}},
  \bibinfo{author}{\bibfnamefont{D.~P.} \bibnamefont{{Schneider}}},
  \bibinfo{author}{\bibfnamefont{D.~J.} \bibnamefont{{Schlegel}}},
  \bibinfo{author}{\bibfnamefont{N.~A.} \bibnamefont{{Bahcall}}},
  \bibinfo{author}{\bibfnamefont{J.~W.} \bibnamefont{{Briggs}}},
  \bibnamefont{et~al.}, \bibinfo{journal}{\apj} \textbf{\bibinfo{volume}{635}},
  \bibinfo{pages}{761} (\bibinfo{year}{2005}).

\end{thebibliography}

\end{document}